\pdfoutput=1
\documentclass[twocolumn,superscriptaddress,aps,preprintnumbers,amsmath,amssymb,prl,nofootinbib]{revtex4}

\usepackage{amsmath}
\usepackage{amssymb}
\usepackage{amsfonts}
\usepackage{graphicx}
\usepackage{xcolor}
\usepackage{xfrac}
\usepackage{comment}
\usepackage{pifont}
\usepackage{physics}
\usepackage{fourier}
\usepackage{hyperref}
\usepackage{bm}
\usepackage{enumitem}
\usepackage{centernot}

\definecolor{rossoferrari}{HTML}{D9073D}
\definecolor{mediumblue}{HTML}{0000CD}
\definecolor{forestgreen}{HTML}{228B22}
\definecolor{desy_blue}{HTML}{009EE2}
\definecolor{desy_orange}{HTML}{FD8800}
\definecolor{light_pink}{rgb}{1,0.4,0.4}
\definecolor{light_blue}{rgb}{0.284602,0.317763,0.963947}
\hypersetup{setpagesize=false,bookmarksnumbered=true,bookmarksopen=true,%
colorlinks=true,linkcolor=light_blue,urlcolor=rossoferrari,citecolor=rossoferrari,linktocpage=false}

\newcommand{\mat}[1]{\bm{\underline{#1}}}
\newcommand{\xmark}{\ding{55}}

\newcommand{\GeV}{\,\mathrm{GeV}}

\begin{document}


\preprint{CERN-TH-2020-196}
\preprint{RESCEU-22/20}
\preprint{DESY 20-202}
\preprint{TU-1112}

\title{Wash-In Leptogenesis}

\author{Valerie Domcke}
\email{valerie.domcke@cern.ch}
\affiliation{Theoretical Physics Department, CERN, 1211 Geneva 23, Switzerland}
\affiliation{Institute of Physics, Laboratory for Particle Physics and Cosmology, EPFL, 1015 Lausanne, Switzerland}

\author{Kohei Kamada}
\email{kohei.kamada@resceu.s.u-tokyo.ac.jp}
\affiliation{Research Center for the Early Universe, The University of Tokyo, Hongo 7-3-1 Bunkyo-ku, Tokyo 113-0033, Japan}

\author{Kyohei Mukaida}
\email{kyohei.mukaida@cern.ch}
\affiliation{Theoretical Physics Department, CERN, 1211 Geneva 23, Switzerland}
\affiliation{Deutsches Elektronen-Synchrotron DESY, 22607 Hamburg, Germany}

\author{Kai Schmitz}
\email{kai.schmitz@cern.ch}
\affiliation{Theoretical Physics Department, CERN, 1211 Geneva 23, Switzerland}

\author{Masaki Yamada}
\email{m.yamada@tohoku.ac.jp}
\affiliation{FRIS, Tohoku University, Sendai, Miyagi 980-8578, Japan}
\affiliation{Department of Physics, Tohoku University, Sendai, Miyagi 980-8578, Japan}

\date{\today}


\begin{abstract}
We present a leptogenesis mechanism based on the standard type-I seesaw model that successfully operates at right-handed-neutrino masses as low as a few 100 TeV.
This mechanism, which we dub \textit{wash-in leptogenesis}, does not require any $CP$ violation in the neutrino sector and can be implemented even in the regime of strong wash-out.
The key idea behind wash-in leptogenesis is to generalize standard freeze-out leptogenesis to a nonminimal cosmological background in which the chemical potentials of all particles not in chemical equilibrium at the temperature of leptogenesis are allowed to take arbitrary values.
This sets the stage for building a plethora of new baryogenesis models where chemical potentials generated at high temperatures are reprocessed to generate a nonvanishing $B\!-\!L$ asymmetry at low temperatures.
As concrete examples, we discuss wash-in leptogenesis after axion inflation and in the context of grand unification.
\end{abstract}


\maketitle


\noindent\textbf{Introduction\,---\,}%
The cosmic imbalance between matter and antimatter~\cite{Aghanim:2018eyx,Zyla:2020zbs} represents clear evidence for new physics beyond the \textit{Standard Model} (SM).
Early attempts to explain the \textit{baryon asymmetry of the Universe} (BAU) related its origin to the $CP$-violating decays of heavy GUT particles in \textit{grand unified theories} (GUTs)~\cite{Yoshimura:1978ex,Dimopoulos:1978kv,Toussaint:1978br,Weinberg:1979bt,Barr:1979ye}.
It was, however, soon realized that electroweak sphaleron processes~\cite{Kuzmin:1985mm} spoil this explanation.
In the early Universe, sphalerons nonperturbatively wash out baryon-plus-lepton number, $B\!+\!L$, which is exactly the linear combination of charges generated during standard GUT baryogenesis.
This observation subsequently led to the proposal of leptogenesis~\cite{Fukugita:1986hr}, which links the BAU to neutrino physics in the type-I seesaw extension of the SM~\cite{Minkowski:1977sc,Yanagida:1979as,Yanagida:1980xy,GellMann:1980vs,Mohapatra:1979ia} and which exploits the fact that sphalerons do not violate baryon-minus-lepton number, $B\!-\!L$.
Indeed, during leptogenesis, the $CP$-violating decays of \textit{right-handed neutrinos} (RHNs) $N_I$ ($I = 1,2,\cdots$) first create a lepton asymmetry (and hence nonzero $B\!-\!L$), which is then converted by the SM interactions in the thermal bath, including sphalerons, to a baryon asymmetry.


Standard thermal leptogenesis requires very large RHN masses, $M_I \gtrsim 10^9\,\textrm{GeV}$, in order to achieve sufficient $CP$ violation during RHN freeze-out~\cite{Davidson:2002qv,Buchmuller:2002rq}. 
This makes it hard to directly probe the RHN sector in experiments and leads to large radiative corrections to the mass of the SM Higgs boson, which aggravates the SM hierarchy problem for RHN masses above the Vissani bound, $ M_I \lesssim 10^7\,\textrm{GeV}$~\cite{Vissani:1997ys,Clarke:2015gwa}. 
In addition, standard leptogenesis is vulnerable to strong asymmetry wash-out, if the RHN Yukawa interactions with the SM lepton--Higgs pairs $\ell_\alpha\,\phi$ are too strong~\cite{Buchmuller:2002jk,Buchmuller:2003gz,Giudice:2003jh,Buchmuller:2004nz}.


In this Letter, we will present a mechanism to generate nonzero $B\!-\!L$ charge in the type-I seesaw model that avoids most of these shortcomings; for alternative routes to low-scale leptogenesis, see~\cite{Akhmedov:1998qx,Pilaftsis:2003gt,Asaka:2005pn,Pilaftsis:2005rv,Hambye:2016sby,Klaric:2020lov,Granelli:2020ysj,Bodeker:2020ghk}.
The key idea behind our proposal is to generalize standard freeze-out leptogenesis to a nonminimal cosmological background in which all conserved charges $C$ at the time of leptogenesis (see Tab.~\ref{tab:charges}) are allowed to take arbitrary values.
In such a background, the \textit{lepton-number-violating} (LNV) RHN interactions then result in a new equilibrium attractor for the chemical potentials in the plasma that generically features nonzero $B-L$, even if $B-L = 0$ initially.
The RHN interactions also actively drive the plasma towards this new attractor solution, which is why we dub our mechanism \textit{wash-in leptogenesis}.


As we will show, wash-in leptogenesis can successfully operate down to RHN masses of a few 100 TeV, \textit{i.e.}, masses shortly above the equilibration temperature of the electron Yukawa interaction~\cite{Bodeker:2019ajh}.
The mechanism therefore allows one to satisfy the Vissani bound; in particular, it is compatible with the \textit{neutrino option}, which denotes the idea that RHNs with masses of a few PeV are responsible for radiatively generating the electroweak scale in the SM~\cite{Brivio:2017dfq,Brivio:2018rzm,Brdar:2019iem,Brivio:2019hrj,Brivio:2020aut}.
Wash-in leptogenesis is also independent of the amount of $CP$ violation in the RHN sector, which liberates it from the Davidson--Ibarra bound, $M_I \gtrsim 10^9\,\textrm{GeV}$; and its success is not jeopardized by large Yukawa couplings.
In fact, in the presence of additional conserved charges, strong asymmetry wash-out turns into efficient asymmetry wash-in.


\begin{table*}[t]
\caption{Decoupling of SM interactions and associated conserved charges $q_C$. 
Yukawa interactions are denoted by $y_i$, weak (strong) sphalerons by WS (SS).
The $\checkmark$ symbol marks efficient interactions.
Hypercharge and the $\Delta_\alpha$ asymmetries are always preserved in the SM.}
\begin{tabular}{c|c||cccccccccccc}
	& $T~[\GeV]$ & $y_e$ & $y_{ds}$ & $y_d$ & $y_s$ & $y_{sb}$ & $y_\mu$ & $y_c$ & $y_\tau$ & $y_b$ & WS & SS & $y_t$ \\
	\hline
	(v) & $\qty(10^5, 10^6)$ & $q_e$ & $\checkmark$ & $\checkmark$ & $\checkmark$ & $\checkmark$ & $\checkmark$ & $\checkmark$ & $\checkmark$ & $\checkmark$ & $\checkmark$ & $\checkmark$ & $\checkmark$ \\
	(iv) & $\qty(10^6, 10^9)$ & $q_e$ & $q_{2 B_1 - B_2 - B_3}$ & $q_{u-d}$ & $\checkmark$ & $\checkmark$ & $\checkmark$ & $\checkmark$ & $\checkmark$ & $\checkmark$ & $\checkmark$ & $\checkmark$ & $\checkmark$ \\
	(iii) & $\qty(10^9, 10^{11-12})$ & $q_e$ & $q_{2 B_1 - B_2 - B_3}$ & $q_{u-d}$ & $q_{d-s}$ & $q_{B_1 - B_2}$ & $q_\mu$ & $\checkmark$ & $\checkmark$ & $\checkmark$ & $\checkmark$ & $\checkmark$ & $\checkmark$ \\
	(ii) & $\qty(10^{11-12}, 10^{13})$ & $q_e$ & $q_{2 B_1 - B_2 - B_3}$ & $q_{u-d}$ & $q_{d-s}$ & $q_{B_1 - B_2}$ & $q_\mu$ & $q_{u-c}$ & $q_\tau$ & $q_{d-b}$ & $q_B$ & $\checkmark$ & $\checkmark$ \\
	(i) & $\qty(10^{13}, 10^{15})$ & $q_e$ & $q_{2 B_1 - B_2 - B_3}$ & $q_{u-d}$ & $q_{d-s}$ & $q_{B_1 - B_2}$ & $q_\mu$ & $q_{u-c}$ & $q_\tau$ & $q_{d-b}$ & $q_B$ & $q_u$ & $\checkmark$
\end{tabular}
\label{tab:charges}
\end{table*}


Our proposal builds on earlier work, which already partly considered some of the ideas presented here~\cite{Campbell:1992jd,Cline:1993vv,Cline:1993bd,Fukugita:2002hu,Fong:2015vna} (see also~\cite{Dick:1999je}).
The essential new elements of our analysis are the following:
(1) We provide a systematic discussion spanning ten orders of magnitude in temperature, $T \in \left(10^5,10^{15}\right)\,\textrm{GeV}$.
In doing so, we account for all possible unconstrained charges in each temperature regime, which allows us to develop a general toolkit for constructing new baryogenesis models; see our main results in Tab.~\ref{tab:toolkit}.
(2) We pay particular attention to flavor.
That is, we allow for an arbitrary flavor composition of the primordial charge asymmetries, and we take into account charged-lepton flavor effects in our analysis of wash-in leptogenesis. 
This especially includes effects related to flavor coherence\,/\,decoherence.
(3) We go beyond LNV two-to-two scattering processes mediated by the dimension-5 Weinberg operator, considering also the ordinary decays and inverse decays of dynamical RHNs.


While wash-in leptogenesis can provide the basis for numerous new baryogenesis models, it does not represent a complete model by itself.
It should rather be regarded as a general mechanism that describes how RHN interactions reprocess primordial charge asymmetries that were generated at higher temperatures.
This includes the intriguing possibility of creating a nonvanishing $B\!-\!L$ asymmetry from $B\!-\!L$-symmetric initial conditions.
But it is agnostic about the \textit{ultraviolet} (UV) physics that is responsible for setting these initial conditions.
This is an advantage, as it allows us to perform a model-independent analysis from a bottom-up perspective.
The remainder of this paper is therefore organized as follows:
First, we will study wash-in leptogenesis in the spirit of an effective field theory that describes the evolution of its input parameters (\textit{i.e.}, the primordial charge asymmetries) from some high-energy matching scale down to low energies.
Then, we will turn to concrete UV completions that illustrate how wash-in leptogenesis can successfully create the BAU, even if $B\!-\!L = 0$ initially.
Specifically, we will consider the generation of nonzero $B\!+\!L$ charge during GUT baryogenesis and axion inflation~\cite{Adshead:2015kza,Adshead:2018oaa,Domcke:2018eki,Domcke:2019mnd}.
A lesson from these examples is that wash-in leptogenesis is able to resurrect baryogenesis scenarios that would otherwise suffer from strong asymmetry wash-out, in a way that is more complex than simply resorting to standard leptogenesis.


\smallskip\noindent\textbf{Wash-in leptogenesis\,---\,}%
We begin by considering a particularly interesting and simple scenario: $N_1$-dominated wash-in leptogenesis at temperatures of a few 100 TeV.
In this temperature regime, all SM interactions are equilibrated\,---\,except for the electron Yukawa interaction, which renders the comoving charge asymmetry of right-handed electrons a classically conserved quantity, $q_e/s = \textrm{const}$, with entropy density $s$.
Its anomalous violation via the chiral plasma instability is negligibly slow for the $q_e/s$ values of interest~\cite{Joyce:1997uy,Kamada:2018tcs,Figueroa:2019jsi}.
At the same time, all charged-lepton flavors $\alpha = e,\mu,\tau$ are fully decohered, which allows us to work with the standard Boltzmann equations for the three lepton flavor asymmetries $\Delta_\alpha = B/3 - L_\alpha$ in the type-I seesaw model~\cite{Pilaftsis:2003gt,Pilaftsis:2005rv},
\begin{equation}
\label{eq:be}
-\left(\partial_t + 3H\right)q_{\Delta_\alpha} = \varepsilon_{1\alpha}\Gamma_1\left(n_{N_1}-n_{N_1}^{\rm eq}\right) - \sum_\beta\gamma_{\alpha\beta}^{\rm w}\,\frac{\mu_{\ell_\beta}+\mu_\phi}{T} \,,
\end{equation}
which is valid in the nonrelativistic regime, $T \lesssim M_1$, where any $N_1$ chemical potential is clearly negligible because of the $N_1$ Majorana mass, $\mu_{N_1} \simeq 0$.
The negative sign on the left-hand side follows from $\Delta_\alpha \supset - L_\alpha$.
The charge asymmetry $q_i$ for a particle species $i$ is defined as the difference of its particle and antiparticle number densities, $q_i = n_i - n_{\bar{\imath}} = g_i\,\mu_i T^2/6$, with chemical potential $\mu_i$ and multiplicity $g_i$, while $q_C = \mu_C T^2/6$ for all conserved charges $C$, with $\mu_C$ in Eq.~\eqref{eq:mu0}.
The first term on the right-hand side of Eq.~\eqref{eq:be} is the standard source term describing the asymmetry production from RHN decays, while the second term is the standard wash-out term, with total wash-out rate per unit volume,
\begin{equation}
\label{eq:gammaw}
\gamma_{\alpha\beta}^{\rm w} = \gamma_{\alpha\beta}^{\rm id} + \sum_\sigma \left[\left(\delta_{\alpha\beta} + \delta_{\sigma\beta}\right)\gamma_{\alpha\sigma}^{\Delta L =2} + \left(\delta_{\alpha\beta} - \delta_{\sigma\beta}\right)\gamma_{\alpha\sigma}^{\Delta L = 0}\right] \,,
\end{equation}
which encompasses RHN inverse decays, $\gamma_{\alpha\beta}^{\rm id} = \gamma_{1\alpha}\,\delta_{\alpha\beta}$ as well as $\Delta L = 2$ and lepton-flavor-violating $\Delta L = 0$ two-to-two scattering processes (see~\cite{Pilaftsis:2003gt,Pilaftsis:2005rv} for more details).


\begin{table*}[t]
\caption{Numerical coefficients $x_C$ that describe the composition of $\mu_{B-L}^{\rm eq} = q_{B-L}^{\rm eq}\,6/T^2$ in terms of the conserved charges $\mu_C = q_C\,6/T^2$ in different temperature regimes; see Eq.~\eqref{eq:qBL}.
The \xmark~symbol marks the absence of the corresponding $\mu_C$ due to an efficient SM interaction.
The second column indicates the active flavors $\ell_\alpha$ with respect to $N_1$ interactions; see the discussion around Eq.~\eqref{eq:lparallel}.
The last column contains $n_{\Delta_\perp}$, which vanishes in the case of $B\!-\!L$-symmetric initial conditions.
$P$ and $P_\tau$ are model-dependent and encode the flavor composition of the primordial $q_{e,\mu,\tau}$ asymmetries with respect to the $N_1$ wash-out direction [see text for examples and Eqs.~\eqref{eq:P}, \eqref{eq:Ptau}].
In this table and throughout the paper, we assume vanishing global hypercharge, $\mu_Y = 0$.
For more details, see the Supplemental Material.}
\begin{tabular}{c|c||c|ccccccccccc|c}
	& $T_{B-L}~[\GeV]$ & Index $\alpha$ &$\mu_e$ & $\mu_{2 B_1 - B_2 - B_3}$ & $\mu_{u-d}$ & $\mu_{d-s}$ & $\mu_{B_1 - B_2}$ & $\mu_\mu$ & $\mu_{u-c}$ & $\mu_\tau$ & $\mu_{d-b}$ & $\mu_B$ & $\mu_u$ & $\mu_{\Delta_\perp}$  \\[.2em]
	\hline
	(v) & $\qty(10^5, 10^6)$ & $e,\mu,\tau$ & $- \frac{3}{10}$ & \xmark & \xmark & \xmark & \xmark & \xmark & \xmark & \xmark & \xmark & \xmark & \xmark & \xmark\\
	(iv) & $\qty(10^6, 10^9)$ & $e,\mu,\tau$  & $- \frac{3}{17}$ & $0$  & $-\frac{7}{17}$ & \xmark & \xmark & \xmark & \xmark & \xmark & \xmark & \xmark & \xmark & \xmark \\
	(iii) & $\qty(10^9, 10^{11-12})$ & $\parallel_\tau,\tau$  
	&$\frac{142 - 225 P_\tau}{247}$ & $0$ & $-\frac{123}{247}$ & $-\frac{82}{247}$ & $\frac{123}{494}$ & $\frac{142 - 225 P_\tau}{247}$ & \xmark & \xmark & \xmark & \xmark & \xmark & $\frac{225}{247}$\\
	(ii) & $\qty(10^{11-12}, 10^{13})$ & $\parallel$  & $\frac{-23 P + 7}{30}$ & $\frac{1}{5}$ & $-\frac{3}{5}$ & $-\frac{1}{6}$ & $-\frac{3}{10}$ & $\frac{-23 P + 7}{30}$ & $\frac{3}{10}$ & $\frac{-23 P + 7}{30}$ & $-\frac{4}{15}$ & $\frac{23}{90}$ & \xmark & $\frac{23}{30}$ \\
	(i) & $\qty(10^{13}, 10^{15})$ & $\parallel$ & $\frac{-3 P + 1}{4}$ & $\frac{1}{6}$ & $-\frac{5}{6}$ & $-\frac{1}{4}$ & $-\frac{1}{4}$ & $\frac{-3 P + 1}{4}$ & $\frac{1}{4}$ & $\frac{-3 P + 1}{4}$ & $-\frac{1}{3}$ & $\frac{1}{6}$ & $\frac{1}{3}$ & $\frac{3}{4}$
\end{tabular}
\label{tab:toolkit}
\end{table*}


Before we are able to solve the coupled system of equations in Eq.~\eqref{eq:be}, we have to specify the relation among the chemical potentials $\mu_{\ell_\alpha}$, $\mu_\phi$ and $\mu_{\Delta_\alpha}$.
In standard leptogenesis, this relation is encoded in the flavor coupling matrix $\left(\mat{C}\right)_{\alpha\beta} = C_{\alpha\beta}$~\cite{Barbieri:1999ma,Abada:2006fw,Nardi:2006fx,Abada:2006ea,Blanchet:2006be,Antusch:2006cw}, whose structure is determined by SM spectator processes~\cite{Buchmuller:2001sr,Garbrecht:2014kda,Garbrecht:2019zaa}. 
The crucial difference between standard leptogenesis and our scenario is that, in a nontrivial chemical background, the standard \textit{linear} relation $\mu_{\ell_\alpha} + \mu_\phi = -\sum_\beta C_{\alpha\beta}\,\mu_{\Delta_\beta}$ turns into an \textit{affine} relation,
\begin{equation}
\label{eq:affine1}
\mu_{\ell_\alpha} + \mu_\phi = \mu_\alpha^0 - \sum_\beta C_{\alpha\beta}\,\mu_{\Delta_\beta} \,,
\end{equation}
where, at temperatures of a few 100 TeV, the translation by the constant shift vector $\mu_\alpha^0$ is solely induced by the conserved chemical potential of the right-handed electrons,
\begin{equation}
\label{eq:affine2}
\begin{pmatrix}
		\mu_{\ell_e}    + \mu_\phi \\
		\mu_{\ell_\mu}  + \mu_\phi \\
		\mu_{\ell_\tau} + \mu_\phi
	\end{pmatrix}
	= 
	\begin{pmatrix}
		-\frac{5}{13} \\
		 \frac{4}{37} \\
		 \frac{4}{37}
	\end{pmatrix} \,\mu_e 
	-
	\begin{pmatrix}
		\frac{6}{13} &  0              & 0             \\
		0            &  \frac{41}{111} & \frac{4}{111} \\
		0            &  \frac{4}{111}  & \frac{41}{111} 
	\end{pmatrix}
	\begin{pmatrix}
		\mu_{\Delta_e}\\
		\mu_{\Delta_\mu}\\
		\mu_{\Delta_\tau}
	\end{pmatrix} \,.
\end{equation}


Eqs.~\eqref{eq:affine1} and \eqref{eq:affine2} follow from analyzing all 16 SM chemical potentials $\mu_i$ ($i = e, \mu, \tau, \ell_e, \ell_\mu, \ell_\tau, u, c, t, d, s, b, Q_1, Q_2, Q_3, \phi$):
In any given temperature regime, the number of linearly independent conserved charges $C$ and the number of SM interactions in equilibrium always add up to 16; see Tab.~\ref{tab:charges}.
This results in 16 constraint equations in each temperature regime that allow one to express the chemical potentials $\mu_i$ of all SM species as linear combinations of the conserved chemical potentials $\mu_C$ ($C = \Delta_\alpha, \cdots$).
In general, we therefore obtain a constant shift vector $\mu_\alpha^0$ in Eq.~\eqref{eq:affine1} of the form
\begin{equation}
\label{eq:mu0}
\mu_\alpha^0 = \sum_{C\neq\Delta_\alpha} S_{\alpha C}\,\mu_C \,, \quad \mu_C = \sum_i n_i^C g_i \mu_i \,,
\end{equation}
with charge vectors $n_i^C$ and multiplicities $g_i$; see \cite{Domcke:2020kcp} for details.
We provide explicit expressions for $n_i^C$, $g_i$, the flavor coupling matrices $C_{\alpha\beta}$, and source matrices $S_{\alpha C}$ in all temperature regimes of interest in the Supplemental Material.


Eqs.~\eqref{eq:be} and \eqref{eq:affine1} tell us that the Boltzmann equations are linear in the lepton flavor asymmetries $\Delta_\alpha$.
This allows us to split $q_{\Delta_\alpha}$ into contributions from thermal and wash-in leptogenesis, respectively, $q_{\Delta_\alpha} = q_{\Delta_\alpha}^{\rm th} + q_{\Delta_\alpha}^{\rm win}$, such that
\begin{equation}
\label{eq:qwin}
\left(\partial_t + 3H\right)q_{\Delta_\alpha}^{\rm win} =  \sum_\beta\Gamma_{\alpha\beta}^{\rm w}\left(q_\beta^0 - \sum_\sigma C_{\beta\sigma}q_{\Delta_\sigma}^{\rm win}\right) \,,
\end{equation}
where $\Gamma_{\alpha\beta}^{\rm w} = 6/T^3\,\gamma_{\alpha\beta}^{\rm w}$.
Eq.~\eqref{eq:qwin} is reminiscent of spontaneous baryogenesis~\cite{Cohen:1987vi,Cohen:1988kt}, specifically, spontaneous leptogenesis~\cite{Kusenko:2014uta,Ibe:2015nfa}, where the rolling of a (pseudo) scalar field $\varphi$ induces effective chemical potentials $\mu_\alpha^0 \propto q_\alpha^0$~\cite{Domcke:2020kcp} (see also \cite{Co:2020xlh,Co:2020jtv}).
The difference between spontaneous leptogenesis and our scenario is that we assume nonzero primordial asymmetries stored in a set of conserved charges, whereas spontaneous leptogenesis involves time-dependent asymmetries\,---\,controlled by the interaction Lagrangian of the field $\varphi$ and not necessarily related to conserved charges\,---\,that are present only when $\varphi$ is in motion.
This requires that LNV processes must be efficient exactly at the time when $\varphi$ is rolling.
In our scenario, such a temporal coincidence is not needed.
Still, it is straightforward to generalize the following analysis to time-dependent charges $q_\alpha^0$~\cite{dkmsy}.


At any given temperature, the total wash-out rate is typically dominated by a single process, such that it factorizes into $\Gamma_{\alpha\beta}^{\rm w} = P_{\alpha\beta}\,\Gamma_{\rm w}$, where the temperature dependence is contained in the flavor-blind wash-out rate $\Gamma_{\rm w}$ and where the matrix $\left(\mat{P}\right)_{\alpha\beta} = P_{\alpha\beta}$ encodes the flavor structure.
In this case, it is then possible to write down an exact solution of Eq.~\eqref{eq:qwin}.
For arbitrary initial conditions $q_{\Delta_\beta}^{\rm ini}$, we find
\begin{equation}
\label{eq:solution}
q_{\Delta_\alpha}^{\rm win} = \sum_\beta \left(\delta_{\alpha\beta} - E_{\alpha\beta}\right) q_{\Delta_\beta}^{\rm eq} + \sum_\beta E_{\alpha\beta}\,q_{\Delta_\beta}^{\rm ini}\,\frac{s}{s^{\rm ini}}\,.
\end{equation}
$q_{\Delta_\alpha}^{\rm eq}$ is the equilibrium attractor in the presence of RHNs,
\begin{equation}
\label{eq:attractor}
q_{\Delta_\alpha}^{\rm eq} = \sum_\beta C_{\alpha\beta}^{-1}\,q_\beta^0 = \sum_\beta \sum_{C\neq\Delta_\alpha} C_{\alpha\beta}^{-1}\,S_{\beta C}\,q_C \,,
\end{equation}
which can also be derived from Eq.~\eqref{eq:affine1} by requiring all RHN interactions to be in equilibrium, $\mu_{\ell_\alpha} + \mu_\phi = \mu_{N_1} = 0$.
The matrix $\left(\mat{E}\right)_{\alpha\beta} = E_{\alpha\beta}$ describes how the RHN interactions actively drive the plasma exponentially close to this solution,
\begin{equation}
\label{eq:E}
\mat{E} = \exp\left(-w\,K_1\,\mat{P\,C}\right) \,,\quad w = \frac{1}{K_1}\int_0^\infty \dd z\: \frac{\Gamma_{\rm w}}{zH} \,,\quad z = \frac{M_1}{T} \,,
\end{equation}
where $K_1$ denotes the standard $N_1$ decay parameter, 
\begin{equation}
K_1 = \frac{\Gamma_1\left(T = 0\right)}{H\left(T = M_1\right)} \,.
\end{equation}
At temperatures of a few 100 TeV, the total wash-out rate is dominated by inverse decays, such that $P_{\alpha\beta} = p_{1\alpha}\,\delta_{\alpha\beta}$ and
\begin{equation}
\mat{E} = \exp\left(-wK_1\mat{C}_1\right) \,,\quad \left(\mat{C}_1\right)_{\alpha\beta} = p_{1\alpha} \,C_{\alpha\beta} \,,\quad p_{1\alpha} = \frac{\Gamma_{1\alpha}}{\Gamma_1} \,,
\end{equation}
where $w \approx 3\pi/4$ assuming Maxwell--Boltzmann statistics for all particles~\cite{Antusch:2010ms}.
For strong wash-in, $K_1 \gg 1$, and a generic flavor structure, $p_{1\alpha} \centernot\ll 1$, all entries of $\mat{E}$ are exponentially suppressed.
The total washed-in $B\!-\!L$ asymmetry then reads
\begin{equation}
\label{eq:qwine}
q_{B-L}^{\rm win} \simeq q_{B-L}^{\rm eq} = \sum_\alpha q_{\Delta_\alpha}^{\rm eq} = -\frac{3}{10}\,q_e \,,
\end{equation}
which also immediately follows from Eq.~\eqref{eq:affine2}.
Any UV mechanism that results in $q_e \neq 0$ at high temperatures thus induces nonzero $B\!-\!L$ at temperatures of a few 100 TeV.


\smallskip\noindent\textbf{Flavor effects\,---\,}%
Next, let us generalize the above discussion to arbitrary temperatures $T \in \left(10^5,10^{15}\right)\,\textrm{GeV}$.
Eqs.~\eqref{eq:be} to \eqref{eq:E}, except for Eq.~\eqref{eq:affine2}, remain valid in this case, the only difference being that the meaning of the flavor index $\alpha$ is now different.
At $T \in \left(10^9,10^{11-12}\right)\,\textrm{GeV}$, electrons and muons propagate as coherent states, which means $\alpha = \parallel_\tau, \tau$, while at temperatures $T \in \left(10^{11-12},10^{15}\right)\,\textrm{GeV}$, all three charged leptons propagate in coherent superpositions, such that $\alpha = \parallel$.
Here, $\ell_\parallel$ represents the coherent single-flavor field that can be created and destroyed by $N_1$ interactions, and $\ell_{\parallel_\tau}$ is the same field after projecting out its $\tau$ component.
Denoting the $N_1$ Yukawa couplings by $h_1^e$, $h_1^\mu$, and $h_1^\tau$, we can write
\begin{equation}
\label{eq:lparallel}
h_\parallel\ell_\parallel = h_1^e\,\ell_e + h_1^\mu\,\ell_\mu + h_1^\tau\,\ell_\tau \,\quad h_{\parallel_\tau} \ell_{\parallel_\tau} = h_1^e\,\ell_e + h_1^\mu\,\ell_\mu \,,
\end{equation}
where $h_\parallel^2 = \left|h_1^e\right|^2 + \left|h_1^\mu\right|^2 + \left|h_1^\tau\right|^2$ and $h_{\parallel_\tau}^2 = \left|h_1^e\right|^2 + \left|h_1^\mu\right|^2$.
Flavor coherence at higher temperatures also implies that some flavor asymmetry $\Delta_\perp$ can escape wash-in leptogenesis,
\begin{equation}
\label{eq:Deltaperp}
\Delta_\perp =
\begin{cases}
B/3 - L_\perp                    & ;\quad T \in \left(10^9,10^{11-12}\right)\,\textrm{GeV}    \\
2B/3 - L_{\perp_1} - L_{\perp_2} & ;\quad T \in \left(10^{11-12},10^{15}\right)\,\textrm{GeV} \\
\end{cases} \,,
\end{equation}
where $\ell_\perp$ is perpendicular to $\ell_\tau$ and $\ell_{\parallel_\tau}$, and $\ell_{\perp_1}$ and $\ell_{\perp_2}$ span the two-dimensional flavor space perpendicular to $\ell_\parallel$.
Making use of these definitions and assuming again strong wash-in and generic RHN couplings, Eq.~\eqref{eq:qwine} now turns into
\begin{equation}
\label{eq:qBL}
q_{B-L}^{\rm eq} = \sum_{C\neq\Delta_\alpha} x_C\,q_C \,, \quad x_C = \delta_{C\Delta_\perp} + \sum_{\alpha,\beta}  C_{\alpha\beta}^{-1}\,S_{\beta C} \,,
\end{equation}
where the numerical coefficients $x_C$ are listed in Tab.~\ref{tab:toolkit}.
This asymmetry remains conserved as soon as the RHN interactions become inefficient at some high temperature $T_{B-L}$~\cite{Buchmuller:2004nz}.
We therefore obtain for the present-day BAU
\begin{equation}
\label{eq:BAU}
\left.\frac{q_B}{s}\right|_{\rm today} = c_{\rm sph}\left.\frac{q_{B-L}^{\rm th} + q_{B-L}^{\rm win}}{s}\right|_{T_{B-L}} \,,
\end{equation}
where $c_{\rm sph} \simeq 12/37$~\cite{Laine:1999wv}.
Note that the standard contribution from thermal leptogenesis may be suppressed because of strong wash-out or insufficient $CP$ violation.


Eq.~\eqref{eq:qBL} and Tab.~\ref{tab:toolkit} are our main results, which serve as a general toolkit to construct new baryogenesis models by implementing the following algorithm:
(1)~Conceive a UV model that leads to primordial chemical potentials $\mu_i$ for some particle species $i$.
(2)~Determine the corresponding conserved charges $\mu_C$.
(3)~Specify the $N_1$ mass and hence relevant temperature scale for leptogenesis, $T_{B-L}$.
(4)~Compute the final BAU according to Eqs.~\eqref{eq:qBL} and \eqref{eq:BAU}.


\smallskip\noindent\textbf{Possible UV completions\,---\,}%
Let us now showcase two possibilities for generating primordial charge asymmetries prior to wash-in leptogenesis.
Both scenarios result in $B\!+\!L \neq 0$ but preserve $B\!-\!L$.
First, we consider $SU(5)$ unification, where the decay of the heavy colored Higgs field $H^\mathrm{c} \subset \bm{5}$ mainly proceeds via the third-generation Yukawa coupling, $H^\mathrm{c} \rightarrow {\bar Q}_3 {\bar Q}_3, t \tau, Q_3 \ell_\tau, {\bar t} {\bar b}$~\cite{Barr:1979ye,Nanopoulos:1979gx,Yildiz:1979gx}.
The production and decay of $H^\mathrm{c}$ bosons after inflation in the $SU(5)$-broken phase (see, \textit{e.g.},~\cite{Kolb:1996jt,Kolb:1998he} for a viable scenario) then results in $\mu_{Q_3} = \mu_{\ell_\tau} = - \mu_\tau = \mu_0$, $\mu_t = -2 \mu_0/3$,  $\mu_b = -\mu_0/3$, or equivalently, $\mu_{B} = -\mu_{2B_1 - B_2 - B_3} = \mu_{\ell_\tau} = - \mu_\tau = 3 \mu_{d-b} = \mu_0$, while all other chemical potentials vanish.
Here, $\mu_0$ is determined by the decay rate, $CP$ violation, and production mechanism of the colored Higgs field. 
This scenario sets the stage for wash-in leptogenesis above the equilibration temperature of the tau Yukawa interaction, $T\gtrsim 10^{11-12}\,\textrm{GeV}$. 
Similarly, one can construct models where extra Higgs scalars also generate primordial asymmetries in the first two fermion generations.
The initial $q_{e,\mu,\tau}$ asymmetries are then encoded in general fields $\bar{e} = c_e e + c_\mu \mu + c_\tau \tau$ or $\bar{e}_\tau = c_e^\tau e + c_\mu^\tau \mu$, such that
\begin{equation}
\label{eq:PPtau}
P = \left|a_e c_e^* + a_\mu c_\mu^* + a_\tau c_\tau^*\right|^2 \,, \quad P_\tau = \left|b_e c_e^{\tau *} + b_\mu c_\mu^{\tau *}\right|^2 
\end{equation}
in Tab.~\ref{tab:toolkit}, where $a_{e,\mu,\tau} = h_1^{e,\mu,\tau}/h_\parallel$ and $b_{e,\mu} = h_1^{e,\mu}/h_{\parallel_\tau}$.


\begin{figure}
\includegraphics[width=0.475\textwidth]{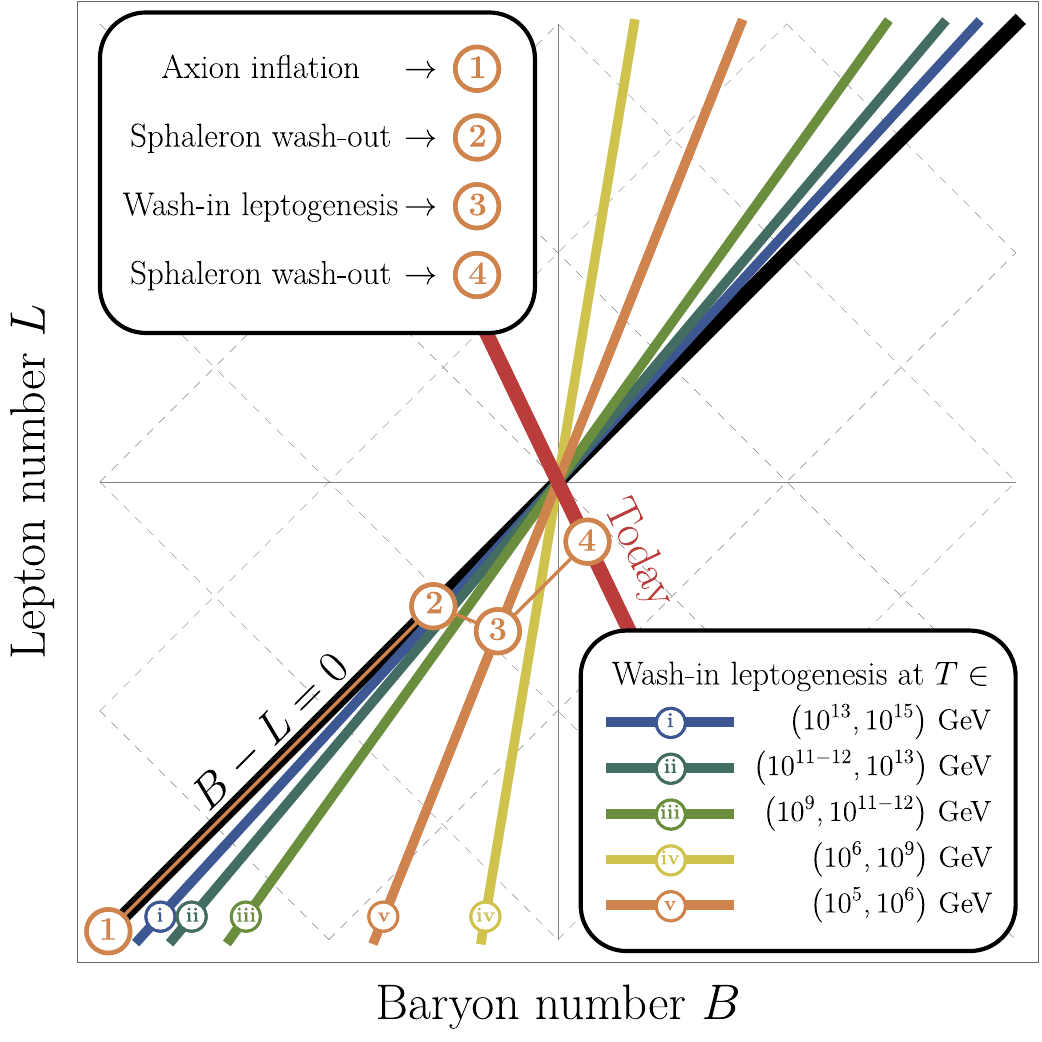}
\caption{Schematic evolution of $B$ and $L$ in arbitrary units after axion inflation.
The colorful straight lines represent the equilibrium attractors of wash-in leptogenesis in different temperature regimes.}
\label{fig:fig1}
\end{figure}


Our second example is axion inflation featuring a coupling of the axion--inflaton field $\varphi$ to the Chern--Simons term of the hypercharge gauge field, $\varphi/\left(4\Lambda\right)\,Y_{\mu\nu}\tilde{Y}^{\mu\nu}$~\cite{Jimenez:2017cdr}.
This coupling sources nonvanishing $\left<Y_{\mu\nu}\tilde{Y}^{\mu\nu}\right>$ during inflation~\cite{Turner:1987bw,Garretson:1992vt,Anber:2006xt}, which induces primordial chemical potentials for all SM fermion species via the SM chiral anomaly~\cite{Adler:1969gk,Bell:1969ts}, $\mu_i/T = \pm 3\,(n_i^{Y})^2 \alpha_Y/\pi\,(h_Y/T^3)_{\rm rh}$~\cite{Domcke:2018eki,Domcke:2019mnd}, with hypercharge fine-structure constant $\alpha_Y$, hypercharges $n_i^Y$, and $\pm$ for left\,/\,right-handed fermions.
$h_Y = \left<\bm{A}_Y\cdot \bm{B}_Y\right>/a^3$ is the physical hypermagnetic helicity density, which is defined in terms of the comoving vector potential $\bm{A}_Y$, comoving flux density $\bm{B}_Y$, and cosmic scale factor $a$.
In the parameter region where $h_{Y}/T^3$ is approximately conserved~\cite{Pouquet:1976zz,Banerjee:2004df,Kahniashvili:2012uj,Domcke:2019mnd}, its value at reheating after inflation dictates the magnitude of the conserved charges in each temperature regime.
For $T \in \left(10^5,10^6\right)\,\textrm{GeV}$, \textit{e.g.}, we have $\mu_e/T = -3\,\alpha_Y/\pi\,(h_Y/T^3)_{\rm rh}$ and hence $\mu_{B-L}/T = 9/10\, \alpha_Y/\pi\,(h_{Y}/T^3)_{\rm rh}$.
Axion inflation with a Hubble rate of $H_{\rm inf} \sim 10^{10}\,\textrm{GeV}$ can therefore readily give rise to the observed baryon asymmetry~\cite{Domcke:2019mnd}.
The evolution of $B$ and $L$ in this scenario is schematically shown in Fig.~\ref{fig:fig1}.
Axion inflation produces all lepton flavors in a symmetric way, meaning $P = 1/3$ and $P_\tau = 1/2$ in Tab.~\ref{tab:toolkit}.


\smallskip\noindent\textbf{Conclusions\,---\,}%
In this Letter, we presented a systematic discussion of \textit{wash-in leptogenesis}, a mechanism to generate nonzero $B\!-\!L$ in the type-I seesaw model.
Our mechanism successfully operates at low RHN masses, strong wash-out, negligible $CP$ violation in RHN decays, and $B\!-\!L$-symmetric initial conditions.
We focused on $N_1$-dominated wash-in leptogenesis; however, the inclusion of heavy-neutrino flavor effects~\cite{Bertuzzo:2010et}, or even the generalization to a density-matrix formalism~\cite{Blanchet:2011xq,Dev:2014laa,Dev:2015wpa}, are straightforward.
Similarly, one may generalize our mechanism to other sources of LNV in the early Universe.
The general concept of wash-in leptogenesis opens the door to a plethora of possibilities.


\medskip\noindent\textit{Acknowledgments\,---\,}%
We thank Apostolos Pilaftsis, Mikhail Shaposhnikov, and Daniele Teresi for helpful comments.
K.\,K.\  was supported by JSPS KAKENHI, Grant-in-Aid for Scientific Research JP19K03842 and Grant-in-Aid for Scientific Research on Innovative Areas 19H04610. 
K.\,M.\ was supported by Deutsche Forschungsgemeinschaft under Germany's Excellence Strategy\,--\,EXC 2121 Quantum Universe\,--\,390833306. 
This project has received funding from the European Union's Horizon 2020 Research and Innovation Programme under grant agreement number 796961, ``AxiBAU'' (K.\,S.).
M.\,Y.\ was supported by the Leading Initiative for Excellent Young Researchers, Ministry of Education, Culture, Sports, Science and Technology (MEXT), Japan, and by JSPS KAKENHI Grants 20H05851, 21K13910, and JP20K22344.


\bibliographystyle{JHEP}
\bibliography{arxiv_2}


\newpage
\onecolumngrid
\newpage


\renewcommand{\thesection}{S\arabic{section}}
\renewcommand{\theequation}{S\arabic{equation}}
\renewcommand{\thefigure}{S\arabic{figure}}
\renewcommand{\thetable}{S\arabic{table}}
\setcounter{equation}{0}
\setcounter{figure}{0}
\setcounter{table}{0}
\setcounter{page}{1}


\begin{center}
\textbf{\Large Supplemental Material: Wash-In Leptogenesis}
\end{center}


In this Supplemental Material, we shall provide an explicit derivation of some of our key results presented in the main text:
(1) the flavor coupling matrices $\left(\mat{C}\right)_{\alpha\beta} = C_{\alpha\beta}$ in Eq.~\eqref{eq:affine1},
(2) the source matrices $\left(\mat{S}\right)_{\alpha C} = S_{\alpha C}$ in Eq.~\eqref{eq:mu0}, and
(3) the numerical coefficients $x_C$ in Eq.~\eqref{eq:qBL} and Tab.~\ref{tab:toolkit}, each respectively evaluated in the five temperature regimes (i) to (v) defined in Tabs.~\ref{tab:charges} and \ref{tab:toolkit}.


\smallskip\noindent\textbf{Chemical equilibrium\,---\,}%
First, let us derive the equilibrium solution for the chemical potentials $\mu_i$ of all SM particle species in the presence of conserved primordial charges.
Our conventions and notation in this subsection will closely follow the discussion in~\cite{Domcke:2020kcp}.
We will regard $\left(\bm{\mu}\right)_i = \mu_i$ as a vector in a 16-dimensional vector space whose index $i$ runs in the following order over the 16 individual fields that make up the particle content of the SM,
\begin{equation}
\label{eq:index}
i = e,\,\mu,\,\tau,\,\ell_e,\,\ell_\mu,\,\ell_\tau,\,u,\,c,\,t,\,d,\,s,\,b,\,Q_1,\,Q_2,\,Q_3,\,\phi\,. 
\end{equation}
At each given temperature, $\bm{\mu}$ is subject to $M + N = 16$ constraint equations, where $M$ is the number of linearly independent SM interactions in chemical equilibrium $I$ and $N$ the number of linearly independent conserved charges $C$,
\begin{equation}
\label{eq:constraints}
\sum_i n_i^I\,\mu_i = \bm{n}^I\cdot\bm{\mu} = 0 \,,\qquad \sum_i n_i^C\,g_i\,\mu_i = \left(\bm{n}^C \circ \bm{g}\right) \cdot \bm{\mu} = \mu_C \,.
\end{equation}
Here, the $\circ$ symbol denotes the entrywise Hadamard product and the quantities $\bm{g}$, $\bm{n}^I$, $\bm{n}^C$, $\mu_C$ are defined as follows:
\begin{itemize}
\item The vector $\bm{g}$ encodes the isospin and color multiplicities of the individual SM fields,
\begin{equation}
\bm{g} = \qty( 1, 1, 1,  2, 2, 2,  3, 3, 3,  3, 3, 3,  6, 6, 6,  4 ) \,.
\end{equation}
\item The $M$ charge vectors $\bm{n}^I$ specify how the SM particles participate in the relevant equilibrated interactions $I$, which can be identified based on the equilibrium temperatures listed in Tab.~\ref{tab:temp}.
For $T \in \left(2.8 \times 10^{13},\,4.7\times10^{15}\right)\,\textrm{GeV}$, \textit{e.g.}, the top-quark Yukawa interaction is the only SM interaction in chemical equilibrium.
The rates of all other interactions are small compared to the Hubble expansion rate at such high temperatures, which prevents the corresponding interactions from reaching chemical equilibrium.
On the other hand, at temperatures below $T \simeq 1.1\times10^5\,\textrm{GeV}$ and above the electroweak phase transition, \textit{all} SM interactions are in chemical equilibrium, including the electron Yukawa interaction, which possess the smallest rate among all SM interactions.
In between these two extremes, more and more SM interactions enter into equilibrium as the temperature decreases and the Hubble expansion rate drops below one interaction rate after another.
The charge vectors $\bm{n}^I$ of these interactions, reaching from the top-quark Yukawa interaction at high temperatures to the electron Yukawa interaction at low temperatures, can be chosen as follows~\cite{Domcke:2020kcp},
\begin{align}
\label{eq:nyt}
\bm{n}^{y_t}    = & \left(0,  0,  0, 0, 0, 0,  0,  0, -1,  0,  0,  0, 0, 0, 1,  1\right) \,, \\
\label{eq:nySS}
\bm{n}^{\rm SS} = & \left(0,  0,  0, 0, 0, 0, -1, -1, -1, -1, -1, -1, 2, 2, 2,  0\right) \,, \\
\bm{n}^{\rm WS} = & \left(0,  0,  0, 1, 1, 1,  0,  0,  0,  0,  0,  0, 3, 3, 3,  0\right) \,, \\
\bm{n}^{y_b}    = & \left(0,  0,  0, 0, 0, 0,  0,  0,  0,  0,  0, -1, 0, 0, 1, -1\right) \,, \\
\bm{n}^{y_\tau} = & \left(0,  0, -1, 0, 0, 1,  0,  0,  0,  0,  0,  0, 0, 0, 0, -1\right) \,, \\
\bm{n}^{y_c}    = & \left(0,  0,  0, 0, 0, 0,  0, -1,  0,  0,  0,  0, 0, 1, 0,  1\right) \,, \\
\bm{n}^{y_\mu}  = & \left(0, -1,  0, 0, 1, 0,  0,  0,  0,  0,  0,  0, 0, 0, 0, -1\right) \,, \\  
\bm{n}^{y_{sb}} = & \left(0,  0,  0, 0, 0, 0,  0,  0,  0,  0, -1,  0, 0, 0, 1, -1\right) \,, \\
\bm{n}^{y_s}    = & \left(0,  0,  0, 0, 0, 0,  0,  0,  0,  0, -1,  0, 0, 1, 0, -1\right) \,, \\
\bm{n}^{y_d}    = & \left(0,  0,  0, 0, 0, 0,  0,  0,  0, -1,  0,  0, 1, 0, 0, -1\right) \,, \\
\bm{n}^{y_{ds}} = & \left(0,  0,  0, 0, 0, 0,  0,  0,  0, -1,  0,  0, 0, 1, 0, -1\right) \,, \\
\bm{n}^{y_e}    = & \left(-1, 0,  0, 1, 0, 0,  0,  0,  0,  0,  0,  0, 0, 0, 0, -1\right) \,.
\label{eq:nye}
\end{align}
The charge vectors of other interactions, such as the up-quark Yukawa interaction or additional flavor-changing Yukawa interactions in the quark sector, can be constructed as linear combinations of these 12 linearly independent vectors.
In the following, it will therefore suffice to work with the vectors listed above, where, among linearly dependent interactions, we always choose the vector corresponding to the interaction that reaches equilibrium at the highest temperature.
We note that the normalization of the vectors $\bm{n}^I$ is meaningless as long as the corresponding interactions are either fully decoupled or fully equilibrated.
This immediately follows from the first relation in Eq.~\eqref{eq:constraints}, which remains invariant under rescalings of $\bm{n}^I$.


\begin{table}
\caption{Equilibration temperatures $T_I$ of all relevant nonperturbative and Yukawa interactions $I$ in the SM; see~\cite{Domcke:2020kcp} for an explicit derivation.
When the temperature of the thermal bath drops below $T_I$ for some $I$, the corresponding process $I$ quickly reaches chemical equilibrium.
Conversely, as long as the temperature exceeds $T_I$, the process $I$ cannot compete with the Hubble expansion of the Universe.
This renders $I$ inefficient and results in the conservation of a global charge that would otherwise be violated by $I$; see Eqs.~\eqref{eq:nie} to \eqref{eq:niu}.}
\begin{tabular}{l||l|l|l}
Nonperturbative processes & Weak sphalerons                                       & Strong sphalerons                                                                                    \\ \hline
                          & $T_{\rm WS} \simeq 2.5 \times 10^{12} \,\mathrm{GeV}$ & $T_{\rm SS} \simeq 2.8 \times 10^{13} \,\mathrm{GeV}$                                                \\ \hline\hline
Yukawa interactions       & First generation                                & Second generation                                  & Third generation                                      \\ \hline
Leptons                   & $T_{y_e} \simeq 1.1 \times 10^5 \,\mathrm{GeV}$ & $T_{y_\mu} \simeq 4.7 \times 10^9 \,\mathrm{GeV}$  & $T_{y_\tau} \simeq 1.3 \times 10^{12} \,\mathrm{GeV}$ \\
Up-type quarks            & $T_{y_u} \simeq 1.0 \times 10^6 \,\mathrm{GeV}$ & $T_{y_c} \simeq 1.2 \times 10^{11} \,\mathrm{GeV}$ & $T_{y_t} \simeq 4.7 \times 10^{15} \,\mathrm{GeV}$    \\
Down-type quarks          & $T_{y_d} \simeq 4.5 \times 10^6 \,\mathrm{GeV}$ & $T_{y_s} \simeq 1.1 \times 10^9 \,\mathrm{GeV}$    & $T_{y_b} \simeq 1.5 \times 10^{12} \,\mathrm{GeV}$
\end{tabular}
\label{tab:temp}
\end{table}


\item The $N$ charge vectors $\bm{n}^C$ contain the charges of all SM particles with respect to the linearly independent global $U(1)_C$ symmetries that are conserved by the $M$ linearly independent SM interactions $I$ in chemical equilibrium.
At the highest temperatures that we are interested in, $T \in \left(2.8 \times 10^{13},\,4.7\times10^{15}\right)\,\textrm{GeV}$, only the top-quark Yukawa interaction is in equilibrium ($M=1$), which results in $N = 15$ linearly independent global charges $C$.
On the other hand, when all $M = 12$ interactions in Eqs.~\eqref{eq:nyt} to \eqref{eq:nye} are equilibrated, only $N = 4$ conserved charges remain: the SM hypercharge $Y$, which is embedded in the SM gauge group, and the three lepton flavor asymmetries $\Delta_e = B/3 - L_e$, $\Delta_\mu = B/3 - L_\mu$, and $\Delta_\tau = B/3 - L_\tau$.
These four charges are conserved by all SM interactions above the electroweak phase transition.
In our analysis, we notably assume $Y = 0$ at all times because of the underlying $U(1)_Y^{\rm local}$ gauge symmetry.
The lepton flavor asymmetries, on the other hand, can obtain nonzero values and play a central role in our proposed mechanism of wash-in leptogenesis (see main text).
The charge vectors of the four global symmetries $U(1)_Y$ and $U(1)_{\Delta_{e,\mu,\tau}}$ read
\begin{align}
\label{eq:Y}\bm{n}^Y                    & = \left(-1, -1, -1, -\sfrac12, -\sfrac12, -\sfrac12, \sfrac23, \sfrac23, \sfrac23, -\sfrac13, -\sfrac13, -\sfrac13, \sfrac16, \sfrac16, \sfrac16, \sfrac12 \right) \,, \\
\label{eq:Deltae}\bm{n}^{\Delta_e}      & = \left(-1,  0,  0,        -1,         0,         0, \sfrac19, \sfrac19, \sfrac19,  \sfrac19,  \sfrac19,  \sfrac19, \sfrac19, \sfrac19, \sfrac19, 0        \right) \,, \\
\label{eq:Deltamu}\bm{n}^{\Delta_\mu}   & = \left( 0, -1,  0,         0,        -1,         0, \sfrac19, \sfrac19, \sfrac19,  \sfrac19,  \sfrac19,  \sfrac19, \sfrac19, \sfrac19, \sfrac19, 0        \right) \,, \\
\label{eq:Deltatau}\bm{n}^{\Delta_\tau} & = \left( 0,  0, -1,         0,         0,        -1, \sfrac19, \sfrac19, \sfrac19,  \sfrac19,  \sfrac19,  \sfrac19, \sfrac19, \sfrac19, \sfrac19, 0        \right) \,.
\end{align}
Any linear combination of conserved charges yields another conserved charge.
This provides us with some freedom in choosing the 11 charges that are successively violated by the the interactions in Eqs.~\eqref{eq:nySS} to \eqref{eq:nye} as the temperature decreases.
A simple and convenient choice of linearly independent charge vectors $\bm{n}^C$ is given by
\begin{align}
\label{eq:nie}
\bm{n}^{y_e}    ~\leftrightarrow~ & \bm{n}^e              = \left(1, 0, 0, 0, 0, 0, 0, 0, 0, 0, 0, 0, 0, 0, 0, 0\right)                                                                      \,, \\
\bm{n}^{y_{ds}} ~\leftrightarrow~ & \bm{n}^{2B_1-B_2-B_3} = \left(0, 0, 0, 0, 0, 0, \sfrac23, -\sfrac13, -\sfrac13, \sfrac23, -\sfrac13, -\sfrac13, \sfrac23, -\sfrac13, -\sfrac13, 0\right) \,, \\
\bm{n}^{y_d}    ~\leftrightarrow~ & \bm{n}^{u-d}          = \left(0, 0, 0, 0, 0, 0, 1, 0, 0, -1, 0, 0, 0, 0, 0, 0\right)                                                                     \,, \\
\bm{n}^{y_s}    ~\leftrightarrow~ & \bm{n}^{d-s}          = \left(0, 0, 0, 0, 0, 0, 0, 0, 0, 1, -1, 0, 0, 0, 0, 0\right)                                                                     \,, \\
\bm{n}^{y_{sb}} ~\leftrightarrow~ & \bm{n}^{B_1-B_2}      = \left(0, 0, 0, 0, 0, 0, \sfrac13, -\sfrac13, 0, \sfrac13, -\sfrac13, 0, \sfrac13, -\sfrac13, 0, 0\right)                         \,, \\
\bm{n}^{y_\mu}  ~\leftrightarrow~ & \bm{n}^\mu            = \left(0, 1, 0, 0, 0, 0, 0, 0, 0, 0, 0, 0, 0, 0, 0, 0\right)                                                                      \,, \\
\bm{n}^{y_c}    ~\leftrightarrow~ & \bm{n}^{u-c}          = \left(0, 0, 0, 0, 0, 0, 1, -1, 0, 0, 0, 0, 0, 0, 0, 0\right)                                                                     \,, \\
\bm{n}^{y_\tau} ~\leftrightarrow~ & \bm{n}^\tau           = \left(0, 0, 1, 0, 0, 0, 0, 0, 0, 0, 0, 0, 0, 0, 0, 0\right)                                                                      \,, \\
\bm{n}^{y_b}    ~\leftrightarrow~ & \bm{n}^{d-b}          = \left(0, 0, 0, 0, 0, 0, 0, 0, 0, 1, 0, -1, 0, 0, 0, 0\right)                                                                     \,, \\
\bm{n}^{\rm WS} ~\leftrightarrow~ & \bm{n}^B              = \left(0, 0, 0, 0, 0, 0, \sfrac13, \sfrac13, \sfrac13, \sfrac13, \sfrac13, \sfrac13, \sfrac13, \sfrac13, \sfrac13, 0\right)       \,, \\
\bm{n}^{\rm SS} ~\leftrightarrow~ & \bm{n}^u              = \left(0, 0, 0, 0, 0, 0, 1, 0, 0, 0, 0, 0, 0, 0, 0, 0\right)                                                                      \,.
\label{eq:niu}
\end{align}
Here, the relations $\bm{n}^I\leftrightarrow\bm{n}^C$ indicate which interactions $I$ violate which global charges $C$.
At $T > T_I$ for some $I$, the charge $C$ is preserved and the corresponding charge vector $\bm{n}^C$ should be used in the \textit{second} relation in Eq.~\eqref{eq:constraints}. 
At $T < T_I$, on the other hand, the interaction $I$ is in chemical equilibrium, the charge $C$ is violated, and the charge vector $\bm{n}^I$ should be used in the \textit{first} relation in Eq.~\eqref{eq:constraints}. 
The equilibration temperatures $T_I$ are listed in Tab.~\ref{tab:temp}.
\item The $N$ chemical potentials $\mu_C$ represent the chemical potentials of all conserved charges $C$ that may, \textit{e.g.}, be determined by a set of primordial chemical potentials $\bm{\mu}^{\rm ini}$ at some high-energy input scale (see main text for examples),
\begin{equation}
\label{eq:muC}
\mu_C = \sum_i n_i^C\,g_i\,\mu_i^{\rm ini} = \left(\bm{n}^C \circ \bm{g}\right) \cdot \bm{\mu}^{\rm ini} \,.
\end{equation}
\end{itemize}


Having introduced these definitions, let us now solve the 16 constraint equations in Eq.~\eqref{eq:constraints} for the chemical potentials $\bm{\mu}$.
To do so, it is convenient to rewrite Eq.~\eqref{eq:constraints} in matrix form,
\begin{equation}
\label{eq:Mmu}
\mat{M}\,\bm{\mu} = \bm{m} \,,\qquad \mat{M} = \begin{pmatrix}\left(\bm{n}^I\right)^{\rm T} \\ \left(\bm{n}^C \circ\bm{g}\right)^{\rm T}\end{pmatrix} \,,\qquad \bm{m} = \begin{pmatrix} 0 \\ \mu_C \end{pmatrix} \,,
\end{equation}
where $\mat{M}$ is a real $\left(M+N\right)\times16 = 16\times16$ matrix whose first $M$ rows contain the transpose of the vectors $\bm{n}^I$ and whose last $N$ rows contain the transpose of the vectors $\bm{n}^C \circ\bm{g}$.
It is now trivial to solve Eq.~\eqref{eq:Mmu} for the chemical potentials $\bm{\mu}$,
\begin{equation}
\label{eq:equilibrium}
\bm{\mu} = \mat{M}^{-1} \bm{m} \,.
\end{equation}
This is the equilibrium solution for the SM chemical potentials in dependence of $\bm{g}$, $\bm{n}^I$, $\bm{n}^C$, $\mu_C$, which we introduced above.


\smallskip\noindent\textbf{Flavor effects\,---\,}%
The Boltzmann equation for the lepton flavor asymmetries $\Delta_\alpha$ involves the chemical potentials of the SM lepton and Higgs doublets, $\mu_{\ell_\alpha}$ and $\mu_\phi$; see the wash-out term on the right-hand side of Eq.~\eqref{eq:be}.
In order to solve the Boltzmann equation, we therefore have to express $\mu_{\ell_\alpha}$ and $\mu_\phi$ in terms of the chemical potentials of the lepton flavor asymmetries, $\mu_{\Delta_\alpha}$.
To derive this relation, we ultimately want to use our result in Eq.~\eqref{eq:equilibrium}.
However, before we are able to do so, we first need to clarify the meaning of the flavor index $\alpha$, which does not necessarily coincide with $e$, $\mu$, and $\tau$ in Eq.~\eqref{eq:index}.


In general, the flavor index $\alpha$ runs over all uncorrelated charged-lepton flavors that actively participate in interactions with $N_1$ RHNs.
These active flavors can be identified based on the RHN Yukawa term in the Lagrangian,
\begin{equation}
\mathcal{L} \supset - \overline{N}_1\,\widetilde{\phi}^\dagger \cdot \left(h_1^e \ell_e + h_1^\mu \ell_\mu + h_1^\tau \ell_\tau\right) + \textrm{H.c.} \,.
\end{equation}
At $T > T_{y_\tau}$, all charged-lepton Yukawa interactions with the SM Higgs field are out of equilibrium.
At such high temperatures, there are hence no SM processes that probe the flavor composition of charged-lepton states, which means that these states remain fully coherent as they propagate through the plasma.
This allows us to define a new charged-lepton field $\ell_\parallel$, with
\begin{equation}
\ell_\parallel = a_e\,\ell_e + a_\mu\,\ell_\mu + a_\tau\,\ell_\tau \,,\qquad a_{e,\mu,\tau} = \frac{h_1^{e,\mu,\tau}}{h_\parallel} \,,\qquad h_\parallel = \left(\left|h_1^e\right|^2 + \left|h_1^\mu\right|^2 + \left|h_1^\tau\right|^2\right)^{1/2} \,,\qquad \mathcal{L} \supset - \overline{N}_1\,\widetilde{\phi}^\dagger \cdot \left(h_\parallel \ell_\parallel\right) + \textrm{H.c.} \,,
\end{equation}
which represents the sole coherent charged-lepton flavor that interacts with $N_1$ at $T > T_{y_\tau}$.
Similarly, we can define $\ell_{\parallel_\tau}$, with
\begin{equation}
\ell_{\parallel_\tau} = b_e\,\ell_e + b_\mu\,\ell_\mu \,,\qquad b_{e,\mu} = \frac{h_1^{e,\mu}}{h_{\parallel_\tau}} \,,\qquad h_{\parallel_\tau} = \left(\left|h_1^e\right|^2 + \left|h_1^\mu\right|^2\right)^{1/2} \,,\qquad \mathcal{L} \supset - \overline{N}_1\,\widetilde{\phi}^\dagger \cdot \left(h_{\parallel_\tau} \ell_{\parallel_\tau} + h_1^\tau \ell_\tau\right) + \textrm{H.c.} \,,
\end{equation}
which represents the coherent charged-lepton flavor in the $e$\,--\,$\mu$ subspace that interacts with $N_1$ at $T_{y_\mu} < T < T_{y_\tau}$, \textit{i.e.}, when only the $\tau$-flavor component of propagating charged-lepton states is measured by interactions in thermal bath.
This means that $\alpha = e,\mu,\tau$ is only true at $T < T_{y_\mu}$.
At $T_{y_\mu} < T < T_{y_\tau}$, we have to work instead with  $\alpha = \parallel_\tau,\tau$, while $T > T_{y_\tau}$, we have  $\alpha = \parallel$.


Despite this temperature-dependent definition of the index $\alpha$, it is still possible to use our result in Eq.~\eqref{eq:equilibrium} to derive a relation among the chemical potentials $\mu_{\ell_\alpha}$, $\mu_\phi$, and $\mu_{\Delta_\alpha}$.
The crucial point is that the SM charged-lepton sector exhibits a $U(3)$ flavor symmetry at $T > T_{y_\tau}$ as well as a $U(2)$ flavor symmetry acting on the $e$\,--\,$\mu$ subspace at $T_{y_\mu} < T < T_{y_\tau}$.
These global flavor symmetries allow us to rotate the $\mu_{e,\mu,\tau}$ and $\mu_{\ell_e,\ell_\mu,\ell_\tau}$ components of the vector $\bm{\mu}$ as well as the corresponding components of the charge vectors in Eqs.~\eqref{eq:Y} to \eqref{eq:niu} to a new basis, consisting of the flavors $\left(\parallel,\perp_1,\perp_2\right)$ at $T > T_{y_\tau}$ and the flavors $\left(\parallel_\tau,\tau,\perp\right)$ at $T_{y_\mu} < T < T_{y_\tau}$, respectively.
Here, $\perp_1$ and $\perp_2$ span the two-dimensional subspace perpendicular to $\parallel$, and $\perp$ is perpendicular to $\parallel_\tau$ and $\tau$, where we assume that $\parallel$ corresponds to a generic flavor direction in the three-dimensional flavor space. 
These basis transformations in the charged-lepton sector leave Eq.~\eqref{eq:equilibrium} invariant, the only difference being that the charged-lepton indices need to be interpreted as $\left(e, \mu,\tau\right)$, $\left(\parallel_\tau,\tau,\perp\right)$, or $\left(\parallel,\perp_1,\perp_2\right)$, depending on the temperature interval.
We are therefore always able to work with Eq.~\eqref{eq:equilibrium} and write down the following decomposition,
\begin{equation}
\label{eq:decomposition}
\mu_{\ell_\alpha} + \mu_\phi = \sum_{C \neq \Delta_\alpha} S_{\alpha C}\,\mu_C - \sum_\beta C_{\alpha\beta}\,\mu_{\Delta_\beta} \,,
\end{equation}
where the first sum runs over all conserved charges except for the lepton flavor asymmetries and where the second sum describes the usual flavor coupling among the charged-lepton flavors.
Explicit expressions for $\mat{S}$ and $\mat{C}$ are given below.


The left-hand side of Eq.~\eqref{eq:decomposition} vanishes when RHN interactions reach chemical equilibrium, $\mu_{\ell_\alpha} + \mu_\phi  = \mu_{N_1} = 0$.
In this case, which we refer to as the strong wash-in regime, it is possible to solve Eq.~\eqref{eq:decomposition} for the lepton flavor asymmetries,
\begin{equation}
\mu_{\Delta_\alpha} = \sum_\beta \sum_{C\neq\Delta_\alpha} C_{\alpha\beta}^{-1}\,S_{\beta C}\,\mu_C \,.
\end{equation}
The total $B\!-\!L$ asymmetry generated during wash-in leptogenesis then follows from summing this expression over all active flavors $\alpha$ and adding the lepton flavor asymmetries in all perpendicular directions in flavor space,
\begin{equation}
\label{eq:sumC}
\mu_{B-L} = \sum_{C\neq\Delta_\alpha} x_C\,\mu_C \,, \quad x_C = \delta_{C\Delta_\perp} + \sum_{\alpha,\beta}  C_{\alpha\beta}^{-1}\,S_{\beta C} \,.
\end{equation}
where $\Delta_\perp = \Delta_{\perp_1} + \Delta_{\perp_2}$ at $T > T_{y_\tau}$, $\Delta_\perp = \Delta_\perp$ at $T_{y_\mu} < T < T_{y_\tau}$, and $\Delta_\perp = 0$ otherwise.
In order to evaluate this expression, it is necessary to know the chemical potentials $\mu_C$ of all conserved charges $C$, which may be generated at some high energy scale; see Eq.~\eqref{eq:muC}.
However, in general, the computation of these chemical potentials is complicated by the fact that the UV process responsible for generating primordial charge asymmetries does not operate in the same charged-lepton flavor basis\,---\,$\left(e, \mu,\tau\right)$, $\left(\parallel_\tau,\tau,\perp\right)$, or $\left(\parallel,\perp_1,\perp_2\right)$\,---\,in which we have performed our calculations up this point.
This pertains in particular to the chemical potentials of the right-handed charged leptons in Eq.~\eqref{eq:sumC}.
In the following, we will therefore show how arbitrary initial conditions in the charged-lepton sector can be mapped onto the three standard chemical potentials $\mu_{e,\mu,\tau}$, irrespective of the temperature regime.
On the one hand, this will result in a general prescription for evaluating Eq.~\eqref{eq:muC} in any model of interest.
On the other hand, it will also allow us to summarize our final results in Tab.~\ref{tab:toolkit} in a unified manner.


First, we consider very high temperatures, $T > T_{y_\tau}$, where the charged-lepton sector enjoys a $U(3)$ symmetry.
In this case, the trace over the chemical potentials of all three right-handed charged leptons is invariant under $U(3)$ transformations,
\begin{equation}
\mu_e + \mu_\mu + \mu_\tau = \mu_\parallel + \mu_\perp \,,
\end{equation}
where $\mu_\perp$ is defined as $\mu_\perp = \mu_{\perp_1} + \mu_{\perp_2}$ at $T > T_{y_\tau}$.
Making use of this relation, we are therefore able to write
\begin{equation}
\label{eq:P}
P = \frac{\mu_\parallel}{\mu_e + \mu_\mu + \mu_\tau} \qquad\Rightarrow\qquad
\mu_\parallel = P \left(\mu_e + \mu_\mu + \mu_\tau\right) \,,\qquad \mu_\perp = \left(1-P\right)\left(\mu_e + \mu_\mu + \mu_\tau\right) \,,
\end{equation}
where the factor $P$ provides a convenient means to relate the chemical potentials in the $\left(\parallel,\perp_1,\perp_2\right)$ basis to the chemical potentials in the $\left(e, \mu,\tau\right)$ basis.
In the next step, we need to specify the initial conditions in the right-handed charged-lepton sector and likewise express them in the $\left(e, \mu,\tau\right)$ flavor basis.
This will allow us to bring together the flavor states defined by the RHN interactions and the flavor states defined by the UV physics in one and the same basis.
Depending on the UV process responsible for generating the primordial charge asymmetries, we may distinguish between the following superpositions of single-particle, two-particle, and three-particle states in the right-handed charged-lepton sector,
\begin{align}
\label{eq:e1}|\bar{e}_{(1)}\rangle & = c_e^* |e\rangle + c_\mu^* |\mu\rangle + c_\tau^* |\tau\rangle \vphantom{\Big[}\,, \\ 
\label{eq:e2}|\bar{e}_{(2)}\rangle & = \frac{1}{2}\:\Big[d_e\left(|\mu\rangle\otimes|\tau\rangle-|\tau\rangle\otimes|\mu\rangle\right) + d_\mu\left(|\tau\rangle\otimes|e\rangle-|e\rangle\otimes|\tau\rangle\right) + d_\tau\left(|e\rangle\otimes|\mu\rangle-|\mu\rangle\otimes|e\rangle\right)\Big]\,, \\
|\bar{e}_{(3)}\rangle & = \frac{1}{6}\:\Big[|e\rangle\otimes|\mu\rangle\otimes|\tau\rangle - |e\rangle\otimes|\tau\rangle\otimes|\mu\rangle - |\mu\rangle\otimes|e\rangle\otimes|\tau\rangle + |\mu\rangle\otimes|\tau\rangle\otimes|e\rangle + |\tau\rangle\otimes|e\rangle\otimes|\mu\rangle - |\tau\rangle\otimes|\mu\rangle\otimes|e\rangle\Big] \,,
\end{align}
which may also be written as
\begin{equation}
|\bar{e}_{(1)}\rangle = c_e^* |e\rangle + c_\mu^* |\mu\rangle + c_\tau^* |\tau\rangle \,,\qquad |\bar{e}_{(2)}\rangle = d_e|\mu\rangle\otimes|\tau\rangle + d_\mu|\tau\rangle\otimes|e\rangle + d_\tau|e\rangle\otimes|\mu\rangle \,,\qquad |\bar{e}_{(3)}\rangle = |e\rangle\otimes|\mu\rangle\otimes|\tau\rangle \,.
\end{equation}
The antisymmetric two- and three-particle states, such as $|e\rangle\otimes|\mu\rangle$ and $|e\rangle\otimes|\mu\rangle\otimes|\tau\rangle$, are product states of pairs and triples of right-handed charged leptons that carry exactly the same quantum numbers except for flavor.
The coefficients $c_{e,\mu,\tau}$ and $d_{e,\mu,\tau}$ are model-dependent and only constrained by the requirement that $|\bar{e}_{(1)}\rangle$ and $|\bar{e}_{(2)}\rangle$ be properly normalized,
\begin{equation}
\big|c_e\big|^2 + \big|c_\mu\big|^2 + \big|c_\tau\big|^2 = 1 \,,\qquad \big|d_e\big|^2 + \big|d_\mu\big|^2 + \big|d_\tau\big|^2 = 1 \,.
\end{equation}
The single-particle state $|\bar{e}_{(1)}\rangle$ can, \textit{e.g.}, be created by UV processes that involve heavy particles decaying into one right-handed charged lepton in the final state. 
As we discuss in the main text, an example of such a process is the decay of heavy colored Higgs fields during GUT baryogenesis.
The fully antisymmetric three-particle state $|\bar{e}_{(3)}\rangle$, on the other hand, can be produced during axion inflation, which treats all charged-lepton flavors on the same footing.
Similarly, the two-particle state $|\bar{e}_{(2)}\rangle$ can be generated by processes that do not distinguish between pairs of charged-lepton flavors.
Note that, for generic values of the complex coefficients $d_{e,\mu,\tau}$, the two-particle state $|\bar{e}_{(2)}\rangle$ typically corresponds to an entangled state.


Based on the above definitions, we are now able to compute the projection factor $P$.
To this end, we first rewrite $P$ as
\begin{equation}
P = \frac{p_\parallel}{p_e + p_\mu + p_\tau} \,,\qquad p_\parallel = \frac{\mu_\parallel}{\mu_{\bar{e}}} \,,\qquad p_{e,\mu,\tau} = \frac{\mu_{e,\mu,\tau}}{\mu_{\bar{e}}} \,,
\end{equation}
where $\mu_{\bar{e}}$ denotes the chemical potential per single flavor that is encoded in the initial right-handed charged-lepton state.
The total chemical potential contained in the initial state $|\bar{e}_{(n)}\rangle$ ($n=1,2,3$) thus corresponds to $n$ times $\mu_{\bar{e}}$. 
The factors $p_\parallel$ and $p_{e,\mu,\tau}$ meanwhile describe the probabilities to find the respective flavor, $\parallel$, $e$, $\mu$, or $\tau$, in the initial state $|\bar{e}_{(n)}\rangle$,
\begin{align}
p_\parallel = \langle\bar{e}_{(n)}|\hat{p}_\parallel|\bar{e}_{(n)}\rangle \,,\qquad p_{e,\mu,\tau} = \langle\bar{e}_{(n)}|\hat{p}_{e,\mu,\tau}|\bar{e}_{(n)}\rangle \,,
\end{align}
where $\hat{p}_\parallel$ and $\hat{p}_{e,\mu,\tau}$ are standard projection operators in the $n$-particle Fock space.
These operators are constructed from single-particle projection operators and the identity operators in the respective orthogonal $\left(n-1\right)$-dimensional subspaces, $\hat{p}_e = |e\rangle\langle e| \otimes \mathbb{1}_{n-1}$ and similarly for all other flavors.
In the $\left(e, \mu,\tau\right)$ and $\left(\parallel,\perp_1,\perp_2\right)$ bases, we can explicitly write
\begin{align}
n = 1 \,: \qquad & \hat{p}_\parallel = |\parallel\rangle\langle\parallel| \,, & \hat{p}_e & = |e\rangle\langle e| \,, \\
n = 2 \,: \qquad & \hat{p}_\parallel = |\parallel\rangle\langle\parallel| \otimes \left(|\perp_1\rangle\langle\perp_1| + |\perp_2\rangle\langle\perp_2|\right) \,, & \hat{p}_e & = |e\rangle\langle e| \otimes \left(|\mu\rangle\langle\mu| + |\tau\rangle\langle\tau|\right) \,, \\
n = 3 \,: \qquad & \hat{p}_\parallel = |\parallel\rangle\langle\parallel| \otimes \left(|\perp_1\rangle\otimes|\perp_2\rangle\langle\perp_1|\otimes\langle\perp_2|\right)  \,, & \hat{p}_e & = |e\rangle\langle e| \otimes \left(|\mu\rangle\otimes|\tau\rangle\langle\mu|\otimes\langle\tau|\right)\,,
\end{align}
and similarly for $\hat{p}_{\mu,\tau}$.
Note that $\hat{p}_\parallel = \hat{p}_{e,\mu,\tau} = |e\rangle\otimes|\mu\rangle\otimes|\tau\rangle\langle e|\otimes\langle\mu|\otimes\langle\tau|$ for $n = 3$.
A straightforward calculation then yields
\begin{align}
n = 1 \,: \qquad & p_\parallel = \left|a_e c_e^* + a_\mu c_\mu^* + a_\tau c_\tau^*\right|^2 \,, & p_{e,\mu,\tau} & = \big|c_{e,\mu,\tau}\big|^2         \,, \vphantom{\Big|} \\
n = 2 \,: \qquad & p_\parallel = 1 - \left|a_e d_e^* + a_\mu d_\mu^* + a_\tau d_\tau^*\right|^2 \,, & p_{e,\mu,\tau} & = 1 - \big|d_{e,\mu,\tau}\big|^2 \,, \vphantom{\Big|} \\
n = 3 \,: \qquad & p_\parallel = 1 \,, & p_{e,\mu,\tau} & = 1                                                                                           \,. \vphantom{\Big|}
\end{align}
In all three cases, we therefore find that the factor $P$ is given by $P = p_\parallel / n$, or more explicitly,
\begin{equation}
\label{eq:Pn}
P =
\begin{cases}
\left|a_e c_e^* + a_\mu c_\mu^* + a_\tau c_\tau^*\right|^2                         & ; \quad n = 1 \vphantom{\Big|} \\
\frac{1}{2}- \frac{1}{2}\left|a_e d_e^* + a_\mu d_\mu^* + a_\tau d_\tau^*\right|^2 & ; \quad n = 2 \vphantom{\Big|} \\
\frac{1}{3}                                                                        & ; \quad n = 3 \vphantom{\Big|}
\end{cases} \,.
\end{equation}


Next, we repeat this analysis in the temperature interval $T_{y_\mu} < T < T_{y_\tau}$, where Eqs.~\eqref{eq:P}, \eqref{eq:e1}, and \eqref{eq:e2} turn into
\begin{align}
\label{eq:Ptau}
P_\tau & = \frac{\mu_{\parallel_\tau}}{\mu_e + \mu_\mu} \qquad\Rightarrow\qquad \mu_{\parallel_\tau} = P_\tau \left(\mu_e + \mu_\mu\right) \,,\qquad \mu_\perp = \left(1-P_\tau\right)\left(\mu_e + \mu_\mu\right) \,, \\
|\bar{e}_{(1)}\rangle_\tau & = c_e^{\tau *} |e\rangle + c_\mu^{\tau *} |\mu\rangle \,, \qquad \big|c_e^{\tau *}\big|^2 + \big|c_\mu^{\tau *}\big|^2 = 1 \,,\\ 
|\bar{e}_{(2)}\rangle_\tau & = \frac{1}{2}\left(|e\rangle\otimes|\mu\rangle-|\mu\rangle\otimes|e\rangle\right) = |e\rangle\otimes|\mu\rangle \,,
\end{align}
which again results in a projection factor $P_\tau = p_{\parallel_\tau}/n$, or equivalently,
\begin{equation}
P_\tau =
\begin{cases}
\left|b_e c_e^{\tau *} + b_\mu c_\mu^{\tau *}\right|^2 & ;\quad n = 1 \\
\frac{1}{2} & ;\quad n = 2
\end{cases} \,.
\end{equation}
Finally, we mention for completeness that one may also consider incoherent combinations of $|\bar{e}_{(1)}\rangle$, $|\bar{e}_{(2)}\rangle$, and $|\bar{e}_{(3)}\rangle$ at $T > T_{y_\tau}$ as well as incoherent combinations of $|\bar{e}_{(1)}\rangle_\tau$ and $|\bar{e}_{(2)}\rangle_\tau$ at $T_{y_\mu} < T < T_{y_\tau}$.
Such situations may, \textit{e.g.}, arise when more than one UV process contributes to the primordial charges or when an initial state $|\bar{e}_{(2)}\rangle$ generated at high temperatures decoheres into an incoherent superposition of $|\bar{e}_{(1)}\rangle_\tau$ and $|\bar{e}_{(2)}\rangle_\tau$ as the temperature decreases.
Generalizing the above computation to these cases is straightforward and simply results in weighted sums of the individual $P$ and $P_\tau$ factors.


\smallskip\noindent\textbf{Final results in five temperature regimes\,---\,}%
We are now able to put everything together and calculate the source and flavor coupling matrices $\mat{S}$ and $\mat{C}$ in Eq.~\eqref{eq:decomposition} as well as the coefficients $x_C$ in Eq.~\eqref{eq:sumC}.
For simplicity, we will not explicitly consider the 11 equilibration temperatures in Tab.~\ref{tab:temp}, but rather work with the five approximate temperature regimes given below.
At the transitions between these regimes, not all SM spectator processes are either fully equilibrated or fully decoupled, which requires a more sophisticated treatment.
\begin{description}
\item[(i) \boldmath{$T \in \left(10^{13},10^{15}\right)\,\textrm{GeV}$}:] Only the top-quark Yukawa interaction is fully equilibrated, which means that we have to compute the matrices $\mat{S}$ and $\mat{C}$ based on Eqs.~\eqref{eq:equilibrium} and \eqref{eq:decomposition} working in the $\left(\parallel,\perp_1,\perp_2\right)$ charged-lepton basis.
In doing so, we always sum over the two perpendicular flavor directions, such that $\Delta_\perp = \Delta_{\perp_1} + \Delta_{\perp_2}$, etc.
This calculation results in
\begin{align}
\mat{S} = \bordermatrix{ & \mu_u   & \mu_B   & \mu_{d-b} & \mu_{u-c} & \mu_{B_1-B_2} & \mu_{d-s} & \mu_{u-d} & \mu_{2B_1-B_2-B_3} & \mu_\parallel & \mu_\perp & \mu_{\Delta_\perp} \cr
                         & \frac29 & \frac19 & -\frac29  & \frac16   & -\frac16      & -\frac16  & -\frac59  & \frac19            & -\frac13      & \frac16   & -\frac16 \cr } \,, \qquad
\mat{C} = \begin{pmatrix}\frac23\end{pmatrix} \,,
\end{align}
where each column of $\mat{S}$ is labeled by the chemical potential of the corresponding conserved charge $C$.
The matrix $\mat{S}$ has one row and the matrix $\mat{C}$ is a $1\times1$ matrix because the index $\alpha$ only runs over a single flavor, $\alpha = \parallel$.
In a second step, we use Eq.~\eqref{eq:P} to express $\mu_\parallel$ and $\mu_\perp$ in the right-handed charged-lepton sector in terms of the standard chemical potentials $\mu_{e,\mu,\tau}$.
Eq.~\eqref{eq:sumC} then leads to the following final expression for $\mu_{B-L}$ in the strong wash-in limit,
\begin{align}
\mu_{B-L} = \frac13\,\mu_u + \frac16\,\mu_B - \frac13\,\mu_{d-b} + \frac14\,\mu_{u-c} - \frac14\,\mu_{B_1-B_2} - \frac14\,\mu_{d-s} - \frac56\,\mu_{u-d} + \frac16\,\mu_{2B_1-B_2-B_3} + \frac{1-3P}{4}\left(\mu_e+\mu_\mu+\mu_\tau\right) + \frac34\,\mu_{\Delta_\perp} \,.
\end{align}
We collect the coefficients on the right-hand side of this expression in the row for temperature regime (i) in Tab.~\ref{tab:toolkit}.
\item[(ii) \boldmath{$T \in \left(10^{11-12},10^{13}\right)\,\textrm{GeV}$}:] The only difference to temperature regime (i) is that now also the strong sphaleron processes are fully equilibrated.
Apart from this, the calculation proceeds exactly as before.
The matrices $\mat{S}$ and $\mat{C}$ read
\begin{equation}
\mat{S} = \bordermatrix{ & \mu_B   & \mu_{d-b}     & \mu_{u-c}    & \mu_{B_1-B_2} & \mu_{d-s}     & \mu_{u-d}     & \mu_{2B_1-B_2-B_3} & \mu_\parallel & \mu_\perp    & \mu_{\Delta_\perp}  \cr
                         & \frac16 & -\frac{4}{23} & \frac{9}{46} & -\frac{9}{46} & -\frac{5}{46} & -\frac{9}{23} & \frac{3}{23}       & -\frac{8}{23} & \frac{7}{46} & -\frac{7}{46} \cr } \,, \qquad
\mat{C} = \begin{pmatrix}\frac{15}{23}\end{pmatrix} \,,
\end{equation}
and the final $B\!-\!L$ asymmetry in the strong wash-in limit is given by [see the row for temperature regime (ii) in Tab.~\ref{tab:toolkit}],
\begin{equation}
\mu_{B-L} = \frac{23}{90} \,\mu_{B} - \frac{4}{15} \,\mu_{d-b} + \frac{3}{10} \,\mu_{u-c} - \frac{3}{10}\,\mu_{B_1-B_2} - \frac{1}{6}\,\mu_{d-s} - \frac{3}{5}\,\mu_{u-d} + \frac{1}{5} \,\mu_{2B_1-B_2-B_3} + \frac{7-23P}{30}\left(\mu_e+\mu_\mu+\mu_\tau\right) + \frac{23}{30}\,\mu_{\Delta_\perp} \,.
\end{equation}
\item[(iii) \boldmath{$T \in \left(10^9,10^{11-12}\right)\,\textrm{GeV}$}:] In this temperature regime, the top-quark Yukawa, strong sphaleron, weak sphaleron, tau Yukawa, bottom-quark Yukawa, and charm-quark Yukawa interactions are equilibrated, which has two important consequences:
(1) We have to switch  from the $\left(\parallel,\perp_1,\perp_2\right)$ basis to the $\left(\parallel_\tau,\tau,\perp\right)$ basis in the charged-lepton sector and (2) the index $\alpha$ now runs over two flavors, $\alpha = \parallel_\tau,\tau$.
Besides that, the matrices $\mat{S}$ and $\mat{C}$ can be computed as before,
\begin{equation}
\mat{S} = \bordermatrix{& \mu_{B_1-B_2}    & \mu_{d-s}       & \mu_{u-d}         & \mu_{2B_1-B_2-B_3} & \mu_{\parallel_\tau} & \mu_\perp      & \mu_{\Delta_\perp} \cr
                        & \frac{123}{2356} & -\frac{41}{589} & -\frac{123}{1178} & 0                  & -\frac{421}{1178}    & \frac{84}{589} & \frac{2}{589} \cr
                        & \frac{42}{589}   & -\frac{56}{589} & - \frac{84}{589}  & 0                  & \frac{86}{589}       & \frac{86}{589} & -\frac{26}{589} \cr } \,,\qquad
\mat{C} = \begin{pmatrix}
\frac{585}{1178} & \frac{26}{589} \\
\frac{26}{589}   & \frac{251}{589}
\end{pmatrix} \,.
\end{equation}
We now use Eq.~\eqref{eq:Ptau} to express $\mu_{\parallel_\tau}$ and $\mu_\perp$ in terms of the standard chemical potentials $\mu_{e,\mu}$, which provides us with the following final $B\!-\!L$ asymmetry in the strong wash-in limit [see the row for temperature regime (iii) in Tab.~\ref{tab:toolkit}],
\begin{equation}
\mu_{B-L} = \frac{123}{494} \,\mu_{B_1-B_2} - \frac{82}{247}\,\mu_{d-s} - \frac{123}{247} \,\mu_{u-d} + \frac{142-225P_\tau}{247}\left(\mu_e + \mu_\mu\right) + \frac{225}{247}\,\mu_{\Delta_\perp} \,. 
\end{equation}
\item[(iv) \boldmath{$T \in \left(10^6,10^9\right)\,\textrm{GeV}$}:] Now all SM interactions except for the Yukawa interactions of the first SM fermion family are equilibrated.
We are therefore able to work in the standard $\left(e,\mu,\tau\right)$ charged-lepton basis, the index $\alpha$ runs over all three flavors, $\alpha = e,\mu,\tau$, and we no longer need to consider any projection factors like $P$ or $P_\tau$.
We find the following expressions for the matrices $\mat{S}$ and $\mat{C}$ and the final $B\!-\!L$ asymmetry in the strong wash-in limit (see also Tab.~\ref{tab:toolkit}),
\begin{equation}
\mat{S} = \bordermatrix{ & \mu_{u-d}       &\mu_{2B_1-B_2-B_3}   & \mu_e            \cr
                         & -\frac{37}{716} & 0                   & -\frac{265}{716} \cr
                         & -\frac{13}{179} & 0                   & \frac{23}{179}   \cr
                         & -\frac{13}{179} & 0                   & \frac{23}{179}   \cr } \,,\qquad
\mat{C} = \begin{pmatrix}
\frac{339}{716} & \frac{3}{179}   & \frac{3}{179}   \\
\frac{3}{179}   & \frac{211}{537} & \frac{32}{537}  \\
\frac{3}{179}   & \frac{32}{537}  & \frac{211}{537}
\end{pmatrix} \,,\qquad
\mu_{B-L} = - \frac{7}{17}\,\mu_{u-d} - \frac{3}{17}\,\mu_{e} \,.
\end{equation}

\item[(v) \boldmath{$T \in \left(10^5,10^6\right)\,\textrm{GeV}$}:] The only difference to temperature regime (iv) is that now all quark Yukawa interactions equilibrated, such that the electron Yukawa interaction is the only SM interaction that has not yet reached chemical equilibrium.
This regime represents our benchmark scenario throughout most of our discussion in the main text.
We find
\begin{equation}
\mat{S} = \bordermatrix{ & \mu_e \cr
                         & -\frac{5}{13} \cr
                         & \frac{4}{37}  \cr
                         & \frac{4}{37}  \cr } \,, \qquad
\mat{C} = \begin{pmatrix}
\frac{6}{13}       & 0 & 0          \\
0 & \frac{41}{111} & \frac{4}{111}  \\
0 & \frac{4}{111}  & \frac{41}{111}
\end{pmatrix} \,,\qquad
\mu_{B-L} = - \frac{3}{10}\,\mu_{e} \,.
\end{equation}

\end{description}


\end{document}